\documentclass[aps,prd,superscriptaddress,reprint,floatfix]{revtex4-1}

\usepackage{amsmath, amsthm, amssymb}
\usepackage{graphicx}

\usepackage{amsmath, amsthm, amssymb}
\usepackage[plainpages=false, colorlinks=true, anchorcolor=blue, linkcolor=blue, citecolor=blue, bookmarks=false]{hyperref}
\usepackage[utf8]{inputenc}
\usepackage{empheq}
\usepackage{graphicx}
\usepackage{siunitx}
\usepackage{xcolor} 
\definecolor{G}{rgb}{0, 0.5, 0}

\usepackage{tikz}
\usetikzlibrary{shapes.geometric, arrows,angles}

\newcommand{\gagg}{g_{a\gamma\gamma}} 

\newcommand{\E}{\mathbf{E}}
\newcommand{\B}{\mathbf{B}}

\begin{document}
\title{Light-shining-through-wall axion detection experiments with a stimulating laser}

\author{K. A. Beyer}
\email[Authors to whom correspondence should be addressed: ]{konstantin.beyer@physics.ox.ac.uk}
\affiliation{Department of Physics, University of Oxford, Parks Road, Oxford OX1 3PU, UK}
\author{G. Marocco}
\email[and:]{giacomo.marocco@physics.ox.ac.uk}
\affiliation{Department of Physics, University of Oxford, Parks Road, Oxford OX1 3PU, UK}
\author{R. Bingham}
\affiliation{Rutherford Appleton Laboratory, Chilton, Didcot OX11 0QX, UK}
\affiliation{Department of Physics, University of Strathclyde, Glasgow G4 0NG, UK}
\author{G. Gregori}
\affiliation{Department of Physics, University of Oxford, Parks Road, Oxford OX1 3PU, UK}

\begin{abstract}
The collision of two real photons can result in the emission of axions. We investigate the performance of a modified light-shining-through-wall (LSW) axion search aiming to overcome the large signal suppression for axion masses $m_a\geq\SI{1}{eV}$. We propose to utilise a third beam to stimulate the reconversion of axions into a measurable signal. We thereby find that with currently available high-power laser facilities we expect bounds at axion masses between $\SI{0.5}{}-\SI{6}{eV}$ reaching $\gagg\geq\SI{e-7}{GeV^{-1}}$. Combining the use of optical lasers with currently operating x-ray free electron lasers, we extend the mass range to $\SI{10}{}-\SI{100}{eV}$. 

\end{abstract}
\maketitle

\section{Introduction}

The Standard Model (SM) is one of the biggest achievements of modern particle physics. While successful in predicting any terrestrial experiment, it is known to be incomplete. It falls short of explaining the CP symmetry of the strong sector and fails to provide explanations for the energy density content of the universe. In fact, only around $5\%$ of the energy density of the universe is in ordinary, baryonic matter, around $26\%$ is in the form of dark matter, which is not contained in the SM. 

One elegant solution to both aforementioned problems makes use of the potential generated by pions after quark confinement. Upon the spontaneous breaking of a new chiral, anomalous $U(1)_\text{PQ}$ symmetry, the CP violating vacuum angle of quantum-chromodynamics (QCD) effectively becomes a dynamic field and runs, in the potential generated by the pions, to the CP-conserving value $\Bar{\theta}\sim 0$. This elegant solution was proposed by Peccei and Quinn in \cite{PhysRevLett.38.1440,PhysRevD.16.1791}. Weinberg and Wilczek pointed out that the spontaneous breaking of the new $U(1)_\text{PQ}$ leads to the appearance of a pseudo Nambu-Goldstone boson, the QCD axion \cite{Weinberg:1977ma,Wilczek:1977pj}. This new particle is a possible candidate to explain the dark matter content of the universe \cite{Preskill:1982cy,Abbott:1982af,Dine:1982ah}.

Generic pseudoscalars also arise abundantly in theory extensions beyond the SM, like in the low energy spectrum of string theory \cite{Witten:1984dg,Arvanitaki:2009fg}. In the following we shall mean by the term axion both, the CP restoring QCD axion and any pseudoscalar particle coupling to electromagnetism with the same 5-dimensional operator
\begin{equation}
    \label{Eq:AxEMcoupl}
    \mathcal{L}_{a\gamma\gamma}=\gagg a\mathbf{E}\cdot\mathbf{B}.
\end{equation}
Here, $\mathbf{E}$ and $\mathbf{B}$ are the electric and magnetic field, respectively and $a$ is the axion field. This interaction is polarisation dependent and thus perfectly suited for laboratory experiments as the coupling can easily be switched off by a simple change of polarisation.

A new, light particle addition to the SM like the axion must be feebly interacting to avoid current detection bounds (see \cite{10.1093/ptep/ptaa104}). Such bounds can be broadly classified into three categories, cosmological, astrophysical and laboratory based. The first two types generically outperform laboratory based searches but suffer from varying model dependence like the underlying assumption that the dark matter content of the universe is fully exhausted by the existence of a single axion. For this reason, laboratory based bounds have been called for \cite{Jaeckel:2006xm}.

Axions are best searched for at the intensity frontier of high power lasers. Axion induced birefringence was searched for by the PVLAS collaboration and it's non-detection placed bounds on the axion parameter space \cite{DellaValle:2015xxa}. Implementing a traditional Sikivie type light shining through wall (LSW) detector \cite{Sikivie:1983ip}, the axion photon coupling $\gagg$ was constrained by multiple groups with the current best bounds set by the QSQAR collaboration \cite{Ballou:2015cka}. We recently proposed a modified experimental approach replacing the static magnetic field of traditional LSW searches by a second laser and thereby avoiding the suppression at large axion masses stemming from the large required momentum transfer \cite{Beyer:2020dag}. The idea behind the proposal of Ref.~\cite{Beyer:2020dag} is a coherent enhancement of the
number of detected photons, $N_\gamma$, that is realised via a standing wave setup. 
However, it can be shown that the setup described in that paper only produces a scaling of $N_\gamma\propto\left|\mathbf{E}\right|^2\left|\mathbf{B}\right|^2\propto N^2$, where $N$ is the number of photons in each of the two lasers used to form the standing wave, instead of the $N^3$ scaling as assumed within the quoted bounds of Ref.~\cite{Beyer:2020dag}. Here we aim to clarify that a $N^3$ enhancement is still possible if the experimental setup is modified by stimulating the photon regeneration process. 

\begin{figure}
\centering
\includegraphics[width=0.45\textwidth]{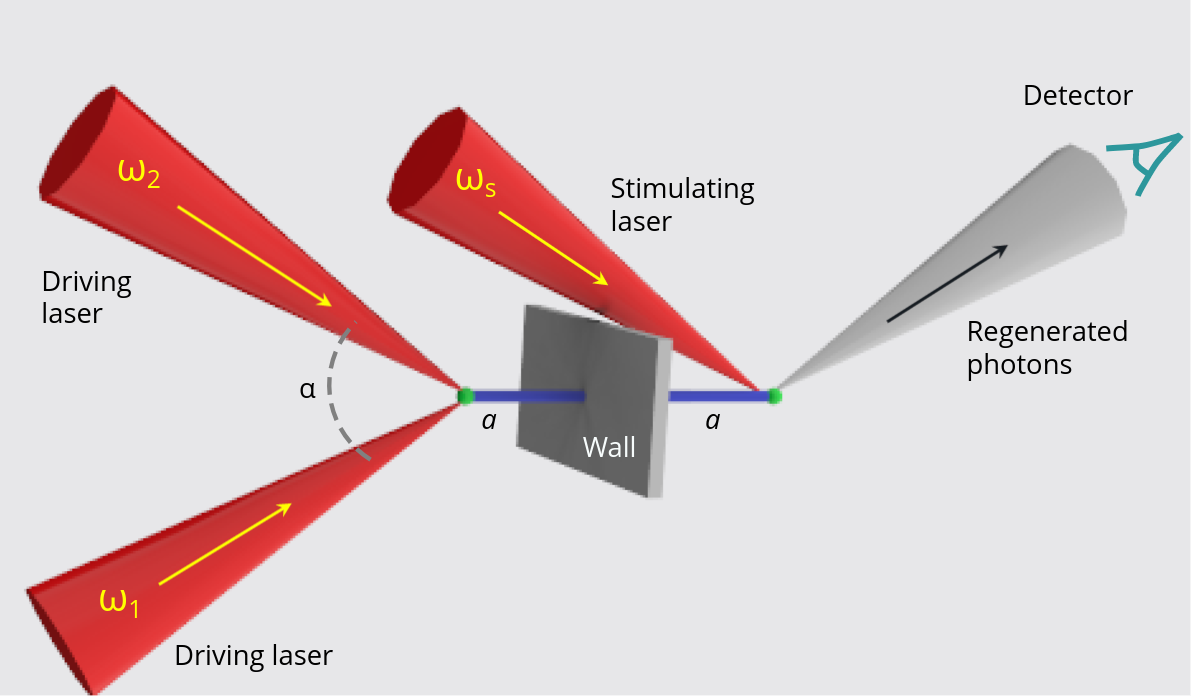}   
\caption{A diagram of the experimental setup. The collision of two lasers results in the production of any hypothetical axions. Such weakly coupled particles pass through a central wall blocking the laser photons from entering the detector region. An appropriately timed third laser facilitates the reconversion into photons behind the wall. Those reconverted photons are measured with a detector.}
    \label{fig:Setup}
\end{figure}

The experimental setup we propose is shown in figure \ref{fig:Setup}. The collision of two lasers produces axions, which, due to their weak coupling, traverse a wall blocking the laser light from penetrating into the detector. We propose to replace the static magnetic field detector of traditional LSW searches by an appropriately timed laser beam, thereby avoiding the large suppression for higher axion masses with a larger required momentum transfer for reconversion. The large photon number in high power laser beams stimulates photon production, further enhancing the signal. The latter was described in \cite{Caputo:2018vmy} for an isotropic photon bath and will be applied to a laser beam in section \ref{Sec:Detection}. Stimulated axion decay can also be used to search for dark matter axions whose decay product produce an echo propagating back to earth when an electromagnetic wave is sent into space \cite{Arza:2019nta,Arza:2021nec}. While we focus on the axion-photon coupling $\gagg$ other schemes investigating the axion's coupling to electrons are investigated in Ref.~\cite{King:2018qbq,Dillon:2018ypt,Dillon:2018ouq}. 

With the above modification, the proposal has similarities to axion searches via light-by-light scattering, in fact it is an on-shell version of it. The distinct advantages lie in the background suppression due to the spatial and temporal separation of the production and reconversion by the interposing wall and macroscopic distance. 
Light-by-light scattering for axion detection was investigated in Ref.~\cite{Shakeri:2020sin}.

The paper is organised as follows, in section \ref{Sec:Production} we review the axion production and calculate the axion field we expect for the aforementioned set-up. The stimulated reconversion of an axion in a laser beam is then investigated in \ref{Sec:Detection} where we find the power in the signal photon field. We finally apply the calculation to our proposed experimental set-up and compare the performance to complementary searches in section \ref{Sec:Signal}.

\section{\label{Sec:Production}Production}

 The presence of an axion $a$ modifies Maxwell's equations \cite{Sikivie:1983ip}
and the resulting wave equations for the fields are
\begin{equation}
    \label{Eq:EWaveEq}
    \left(\partial_t^2-\nabla^2\right)\mathbf{E}=\gagg\left[\partial_t\left(\mathbf{B}\partial_ta-\mathbf{E}\times\nabla a\right)-\nabla\left[\left(\nabla a\right)\cdot\mathbf{B}\right]\right]
\end{equation}
and
\begin{equation}
    \label{Eq:BWaveEq}
    \left(\partial_t^2-\nabla^2\right)\mathbf{B}=\gagg
    \nabla\times\left(\mathbf{E}\times\nabla a-\mathbf{B}\partial_ta\right).
\end{equation}
The axion field obeys the Klein-Gordon equation
\begin{equation}
    \label{Eq:KleinGordon}
    \left(\partial_t^2-\nabla^2+m_a^2\right)a=-\gagg\mathbf{E}\cdot\mathbf{B},
\end{equation}
where $m_a$ is the axion mass.
The electric and magnetic fields are produced by two linearly polarised laser beams colliding at an angle $\alpha$. If the pulse length $T$ is much greater than the central frequency $\omega_j$, $j=1,2$, of each laser beam, then the electric and magnetic fields, $\E_j$ and $\B_j$, respectively, may be treated as a single plane-wave. These may then be decomposed as
\begin{align}
    \label{Eq:InitialBeams}
    \mathbf{E}_j&=\frac{1}{2}\left(\boldsymbol{\mathcal{E}}_j e^{i\omega_j t-i\mathbf{k}_j\cdot\mathbf{x}}+c.c.\right),\nonumber\\ \mathbf{B}_j&=\frac{1}{2}\left(\boldsymbol{\mathcal{B}}_j e^{i\omega_j t-i\mathbf{k}_j\cdot\mathbf{x}}+c.c.\right),
\end{align}
while the axion sourced by these fields is
\begin{equation}
    \label{Eq:AxionFieldDef}
    a=\frac{1}{2}\left[\Tilde{a}(\mathbf{x})e^{i\omega_a t}+ c.c.\right],
\end{equation}
with
\begin{equation}
    \label{Eq:ConversionAxion}
    \left(\omega_a^2+\nabla^2-m_a^2\right)\Tilde{a}(\mathbf{x})e^{i\omega_a t}=\frac{\gagg}{2}\mathcal{F}e^{i(\omega_1+\omega_2)t-i(\mathbf{k}_1+\mathbf{k}_2)\cdot\mathbf{x}}
\end{equation}
where $\mathcal{F}=\left(\boldsymbol{\mathcal{E}}_1\cdot\boldsymbol{\mathcal{B}}_2+\boldsymbol{\mathcal{E}}_2\cdot\boldsymbol{\mathcal{B}}_1\right)$ and $\mathbf{x}$ is the position vector. We will in the following adopt a coordinate system centred on the axion production region. The long laser pulse length also fixes the axion energy $\omega_a=\omega_1+\omega_2$ and we define $\mathbf{k}_a=\mathbf{k}_1+\mathbf{k}_2$, whose magnitude is 
\begin{equation}
    \label{Eq:AngleMass}
    \lvert \mathbf{k}_a\rvert=\sqrt{(\omega_1+\omega_2)^2-4\omega_1\omega_2\sin^2\frac{\alpha}{2}}\equiv\sqrt{\omega_a^2-m_a^2}.
\end{equation}
Hence, we see that the collision angle $\alpha$ sets the axion mass the set-up tests
\begin{equation}
    \label{Eq:AxMass}
    m_a=\sqrt{4\omega_1\omega_2\sin^2\frac{\alpha}{2}}.
\end{equation}

The fundamental solution to the axion equation is
\begin{equation}
    \label{Eq:AxionFundamentalSol}
    G(\mathbf{x})=\int\frac{d^3k}{(2\pi)^3}\frac{e^{i\mathbf{k}\cdot\mathbf{x}}}{-\omega_a^2+\mathbf{k}^2+m_a^2}=\frac{e^{-ik_a|\mathbf{x}|}}{4\pi|\mathbf{x}|},
\end{equation}
where we neglected the advanced solution and only keep the retarded one. The axion field is then obtained via an integration over the beam overlap region, $V$,
\begin{align}
    \label{Eq:AxionSolDef}
    \Tilde{a}(\mathbf{x})&\equiv-\frac{\gagg}{2}\mathcal{F}\int d^3y G(\mathbf{x}-\mathbf{y})e^{-i\mathbf{k}_a\cdot\mathbf{y}}\nonumber\\
    &\simeq -\frac{\gagg}{8\pi}\mathcal{F}\frac{e^{-ik_a|\mathbf{x}|}}{|\mathbf{x}|}\int_V d^3y e^{i k_a\left(\hat{x}-\hat{k}_a-\frac{\mathbf{y}}{2|\mathbf{x}|}+\frac{\hat{x}\cdot\mathbf{y}}{2|\mathbf{x}|}\hat{x}\right)\cdot\mathbf{y}},
\end{align}
which applies in the limit where we evaluate the field far from the overlap volume, that is $\lvert \mathbf{x} \rvert \gg\lvert\mathbf{y}\rvert$ and $\lvert \mathbf{x} \rvert\gg \sqrt{k_a \lvert \mathbf{y}\rvert^3} $.

We can make further simplifications if we consider the direction along the axion momentum $\hat{x}-\hat{k}_a\simeq 0$, where $\mathbf{x}/|\mathbf{x}|\equiv\hat{x}$. Treating the overlap $V$ as a cube of sidelength $\ell\ll |\mathbf{x}|$, the integral becomes
\begin{align}
    \label{Eq:ApproxProduction2}
    \frac{1}{V}&\int_V d^3y e^{-i k_a\left(\frac{\mathbf{y}}{2|\mathbf{x}|}-\frac{\hat{x}\cdot\mathbf{y}}{2|\mathbf{x}|}\hat{x}\right)\cdot\mathbf{y}}\nonumber\\
   &= \left\{\frac{\sqrt{\pi}(1-i)}{\sqrt{\frac{k_a \ell^2}{|\mathbf{x}|}}}\text{Erf}\left[\frac{(1+i)}{4}\sqrt{\frac{k_a \ell^2}{|\mathbf{x}|}}\right] \right\}^2.
\end{align}
As we increase the spot size of the two incoming lasers, therefore increasing $\ell$ and the interaction volume, the axion field amplitude grows linear in volume as long as $\ell^2 < d/k_a$ where we define $d$ the distance to the reconversion region. Increasing the spot size further will only produce a growth linear in $\ell$. In fact the situation is worse because the above scaling is strictly only true when keeping the laser fields $\boldsymbol{\mathcal{E}}_j$ constant. In a real laser system of course the energy is constant and therefore the fields scale like $|\boldsymbol{\mathcal{E}}_j|^2\propto \ell^{-2}$ resulting in an optimal spot size set by the ratio of separation to axion momentum. In this limit we may approximate \eqref{Eq:ApproxProduction2} by $V$ thus resulting in an axion field given by
\begin{equation}
    \label{Eq:AxionSolDistance}
    \Tilde{a}(\mathbf{x})e^{i\omega_a t}=-\frac{\gagg}{8\pi}V\mathcal{F}e^{i\omega_a t}\frac{e^{-i\sqrt{\omega_a^2-m_a^2}|\mathbf{x}|}}{|\mathbf{x}|},
\end{equation}
at large distances along $\hat{k}_a$.

Using this result, we can also calculate the gradient along the observation direction at the position of axion reconversion into photons, that is
\begin{equation}
    \label{Eq:AxionSolGradient}
    \nabla\Tilde{a}(d)=-\mathbf{k}_a\frac{1+ik_a d}{k_a d}\Tilde{a}(d),
\end{equation}
which we will use later. 

\section{\label{Sec:Detection}Axion Reconversion}
 The produced axions must now be reconverted into photons to leave a detectable signal. Here, we propose to place an opaque wall in the way of the source (drive) lasers, through which, instead, all axions can pass through. These are reconverted by a third laser beam via stimulated axion decay. For simplicity, we choose the stimulating beam to be a copy of either one of the initial beams incident at the same angle on the other side of the wall, see figure \ref{fig:Setup}. 

The calculation of the signal power proceeds in much the same way as the previous calculation. The only difference is in the perturbation theory of the axion reconversion, in that now we start with an axion field and only one laser, labelled by $s$. We parametrize the signal field as
\begin{equation}
    \label{Eq:SignalField}
    \mathbf{E}=\frac{1}{2}\left(\Tilde{\mathbf{E}}(\mathbf{x}) e^{i\omega t}+c.c.\right), \qquad \mathbf{B}=\frac{1}{2}\left(\Tilde{\mathbf{B}}(\mathbf{x}) e^{i\omega t}+c.c.\right).
\end{equation}
The equations describing the axion-sourced electric field are
\begin{widetext}
\begin{equation}
    \label{Eq:ReconversionElectricFieldEquation}
    \left(-\omega^2-\nabla^2\right)\Tilde{\mathbf{E}}(\mathbf{x})e^{i\omega t} = -\frac{\gagg}{2}\left\{i(\omega_a-\omega_s)\left(\boldsymbol{\mathcal{E}}^*_s \times\nabla\Tilde{a}-i\omega_a\boldsymbol{\mathcal{B}}^*_s\Tilde{a}\right)e^{i\mathbf{k}_s\cdot\mathbf{x}}+\nabla\left[\left(\nabla\Tilde{a}\right)\cdot\boldsymbol{\mathcal{B}}^*_se^{i\mathbf{k}_s\cdot\mathbf{x}}\right]\right\}e^{i(\omega_a-\omega_s)t},
\end{equation}
and magnetic field
\begin{equation}
\label{Eq:ReconversionMagenticFieldEquation}
    \left(-\omega^2-\nabla^2\right)\Tilde{\mathbf{B}}(\mathbf{x})e^{i\omega t} = \frac{\gagg}{2}\nabla\times\left(\boldsymbol{\mathcal{E}}^*_s e^{i\mathbf{k}_s\cdot\mathbf{x}}\times\nabla\Tilde{a}-i\omega\boldsymbol{\mathcal{B}}^*_se^{i\mathbf{k}_s\cdot\mathbf{x}}\Tilde{a}\right)e^{i(\omega_a-\omega_s)t}.
\end{equation}
\end{widetext}
The signal photon energy is $\omega=\omega_a-\omega_s$. Let us start with the electric field and calculate the source
from the axion field \eqref{Eq:AxionSolDistance} at the reconversion area which is a large distance $d$ away from the conversion in the direction of the axion momentum. The source density generating the electric field on the right hand side of \eqref{Eq:ReconversionElectricFieldEquation} is then
\begin{equation}
    \label{Eq:SourceReconversion}
    \mathbf{j}(\mathbf{x})=-\gagg\mathbf{j}_0\Tilde{a}(\mathbf{x}) e^{i\mathbf{k}_s\cdot(\mathbf{x})},
\end{equation}
with
\begin{align}
    \label{Eq:SourceReconversionConst}
    \mathbf{j}_0=& i\omega \left(\left(\boldsymbol{\mathcal{E}}^*_s\times\mathbf{k}_a\right)\frac{1+i k_a d}{k_a d}-i\omega_a \boldsymbol{\mathcal{B}}_s^*\right) \nonumber\\
    &+\mathbf{k}_a\left(\boldsymbol{\mathcal{B}}^*_s\cdot\mathbf{k}_a\right)\left(\frac{1+i k_a d}{k_a d}\right)^2-i\mathbf{k}_s \left(\boldsymbol{\mathcal{B}}^*_s\cdot\mathbf{k}_a\right)\frac{1+i k_a d}{k_a d}.
\end{align}
We find the electric field from the fundamental solution \eqref{Eq:AxionFundamentalSol} in analogy to before
\begin{equation}
    \label{Eq:ElectricFieldSolDef}
    \Tilde{\mathbf{E}}(\mathbf{x})\simeq-\gagg\frac{e^{-i\omega |\mathbf{x}|}}{8\pi|\mathbf{x}|}\mathbf{j}_0\Tilde{a}(d)\int_{V'}d^3 ye^{-i\omega(\hat{k}-\hat{x})\cdot\mathbf{y}},
\end{equation}
where $V'$ is the volume of the reconversion region, and, again, we evaluate the field in the far-field limit $|\mathbf{x}|\gg k \ell'^2$ with $\ell'$ the sidelength of the reconversion volume, approximated by a cube, and we take the envelope of the source constant over $V'$. This time however we wish to maximise the solid angle over which we collect the signal photons, hence we may no longer limit ourselves to a direction parallel $\hat{k}$. 

To estimate the signal power we are interested in the intensity in the electromagnetic field a detector at distance $D$ covering a solid angle $d^2\Omega$. In the presence of an axion field a non-zero scalar potential $\Phi$ is generated via $\nabla^2\Phi=-\gagg(\nabla a)\cdot\mathbf{B}$ resulting in an electric field component parallel to the gradient. Such a field does not propagate in vacuum and will not reach the detector. We may either choose the stimulating laser beam such that $\gagg(\nabla a)\cdot\mathbf{B}=0$ and hence restore the gauge freedom to set $\Phi=0$, at least to first order in $\gagg$, or we must limit the detected power to the electric field component orthogonal to the photon momentum $\hat{k}$. The power reaching the detector at distance $D$ is then
\begin{equation}
    \label{Eq:Power}
    \mathcal{P}=\int d\vartheta d\varphi \sin(\vartheta) D^2 \left|\Tilde{\mathbf{E}}(D,\vartheta,\varphi)_\perp\right|^2\cos^2\left(\omega t\right).
\end{equation}
The integrand is highly peaked around the photon momentum justifying an integration over the whole sphere as long as our detector is large enough. We thence find, to leading order in $(\ell'\omega)^{-1}$, for the square of the volume integral in equation \eqref{Eq:ElectricFieldSolDef}:
\begin{widetext}
\begin{equation}
    \label{Eq:PowerIntegral}
    \frac{64}{\omega^6}\int d\vartheta d\varphi \sin(\vartheta) \frac{\sin^2\left\{\frac{\ell'\omega}{2}[1-\sin(\theta)\cos(\varphi)]\right\}}{(1-\sin(\theta)\cos(\varphi))^2}\frac{\sin^2\left[\frac{\ell'\omega}{2}\sin(\theta)\sin(\varphi)\right]}{(\sin(\theta)\sin(\varphi))^2}\frac{\sin^2\left[\frac{\ell'\omega}{2}\cos(\theta)\right]}{\cos^2(\theta)}\simeq\frac{4\pi^2\ell'^6}{(\ell'\omega)^2}
\end{equation}
\end{widetext}
where for the exact form of the angular dependence, we assumed the interaction volume to be oriented such that $\hat{k}$ is a unit vector pointing towards one of the faces of the cube $V'$. Any other orientation should not change the solution significantly.

We define the geometry of the setup for two laser beams of equal frequency 
\begin{align}
    \label{Eq:GeometryDef}
    \frac{\boldsymbol{\mathcal{E}}_1}{|\boldsymbol{\mathcal{E}}_1|}&=\begin{pmatrix}0 \\1 \\0\end{pmatrix}, &\frac{\boldsymbol{\mathcal{B}}_1}{|\boldsymbol{\mathcal{B}}_1|}&=\begin{pmatrix}-\cos\frac{\alpha}{2}\\0 \\-\sin\frac{\alpha}{2} \end{pmatrix},  &\frac{\mathbf{k}_1}{\omega_1}&=\begin{pmatrix}-\sin\frac{\alpha}{2} \\0 \\\cos\frac{\alpha}{2}\end{pmatrix}\\
    \frac{\boldsymbol{\mathcal{B}}_2}{|\boldsymbol{\mathcal{B}}_2|}&=\begin{pmatrix}0 \\1 \\0\end{pmatrix}, &\frac{\boldsymbol{\mathcal{E}}_2}{|\boldsymbol{\mathcal{E}}_2|}&=\begin{pmatrix}\cos\frac{\alpha}{2} \\0 \\-\sin\frac{\alpha}{2}\end{pmatrix}, &\frac{\mathbf{k}_2}{\omega_2}&=\begin{pmatrix}\sin\frac{\alpha}{2} \\0 \\\cos\frac{\alpha}{2}\end{pmatrix}\\
    \boldsymbol{\mathcal{E}}_s&=\boldsymbol{\mathcal{E}}_2,  &\boldsymbol{\mathcal{B}}_s&=\boldsymbol{\mathcal{B}}_2,  &\mathbf{k}_s&=\mathbf{k}_2,
\end{align}
where $\alpha$ is the angle between the two drive beams (see Figure 1).
For a generalisation to beams with different frequencies see Appendix \ref{App-Assym}.

The incoming beams are focused such that the beams are cubes of side $\ell$, hence the laser energy contained in the matching interaction volume $E_j=\int\mathcal{P}_jd\tau=|\boldsymbol{\mathcal{E}}_j|^2 \ell^3/2$ is simply the laser energy per pulse. This results in the energy of the signal field to be
\begin{equation}
    \label{Eq:PowerOutput}
    E=\frac{\gagg^4}{64\pi^2} \frac{\ell^2}{d^2}\omega_a^2E_1E_2^2\sin^4\frac{\alpha}{2}\left(1-\frac{k_a}{\omega_a}\cos\frac{\alpha}{2}\right)^2
\end{equation}
where we set $\ell=\ell'$ because the pulselength of the stimulating laser should not be longer than the initial lasers and for simplicity we take it to be a cube again. The dependence on the scattering angle $\alpha$ can be rewritten as an axion mass dependence through \eqref{Eq:AxMass}, resulting, in the case of $\omega_1=\omega_2$, in
\begin{equation}
    \label{Eq:PowerOutputMass}
    E=\frac{\gagg^4}{64\pi^2} \frac{\ell^2}{d^2}m_a^2E_1E_2^2\left(\frac{m_a}{\omega_a}\right)^6.
\end{equation}
In the general case with different frequency beams the dependence may be more complicated, see Appendix \ref{App-Assym}.

Note, we chose the stimulating beam to be the same as beam $2$, such that $\boldsymbol{\mathcal{B}}_s^*\cdot\mathbf{k}_a=0$, simplifying the expression for $\mathbf{j}_0$. Performing the full calculation for the other choice of stimulating beam results in the same bounds, thus, justifying this simplifying assumption.

\section{\label{Sec:Signal}Projected Bounds}
\begin{figure}
    \centering
    \includegraphics[width=0.45\textwidth]{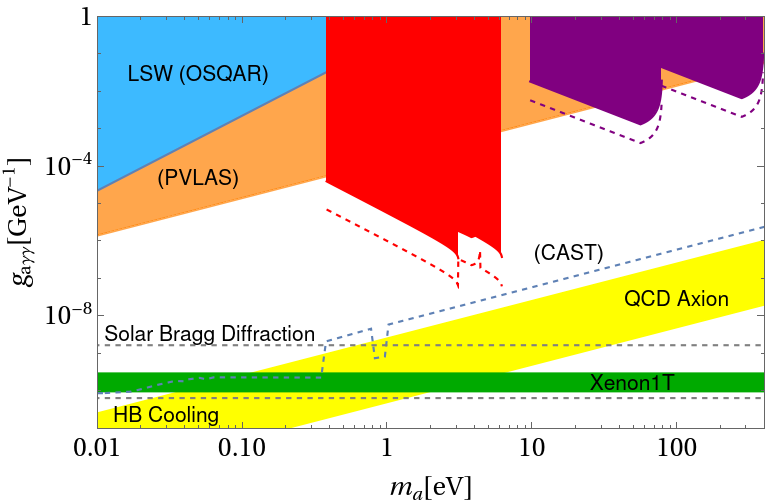}
    \caption{Exclusion plot for axion parameter space. The light blue region shows existing bounds from the OSQAR experiment \cite{Ballou:2015cka}; the orange region is excluded by PVLAS \cite{DellaValle:2015xxa}; the dashed blue line depicts CAST constraints \cite{anastassopoulos2017new}; the lower horizontal dashed line comes form stellar cooling lifetimes \cite{ayala2014revisiting} and the upper from solar Bragg diffraction experiments \cite{bernabei2001search}. The green region shows the Xenon1T anomaly interpreted as QCD axion signal \cite{Aprile:2020tmw,Dent:2020jhf}. The red region on the left indicates the reach of the set-up described in the main text using three optical lasers. We included the mass interval which can be probed when considering frequency doubled beams. The dashed red line indicates the improvement for a $\SI{15}{kJ}$ laser. The purple region on the right shows the projected bounds for the collision of an optical $\SI{1.5}{kJ}$ laser and an X-FEL like the european X-FEL. The bounds extend from $\omega=\SI{1}{keV}$, on the left, to $\omega=\SI{25}{keV}$ on the right and again, the dashed line is a projection to $\SI{15}{kJ}$ optical laser energy. The QCD axion region, shown in yellow, indicates particular theoretical predictions for where the axion might be, given dark matter abundances \cite{di2017redefining}.}
    \label{fig:Reach}
\end{figure}

To assess the performance of the above proposal we will evaluate the projected bounds utilising the Aton 4 laser at the Extreme Light Infrastructure (ELI) beamlines. This laser system operates at optical frequencies $\omega_j=\SI{1.55}{eV}$ ($j=1,2$) with $E_j=\SI{1.5}{kJ}$ energy per pulse and has pulse lengths of $\SI{150}{fs}$ up to $\tau=\SI{1}{ns}$. The optimal pulse duration was discussed earlier and turned out to be $\tau=\sqrt{d/k_a}$. The number of signal photons incident on the detector can simply be obtained from the energy equation \eqref{Eq:PowerOutput} as $N_\gamma=E/\omega$. We will in the following assume single photon counting is possible using a transition edge detector similar to the one designed for the ALPS II experiment \cite{bastidon2016quantum,spector2016alps} and exploiting the coincidence timing of signal and incoming lasers to discriminate background. The Aton 4 laser has a repetition rate of $\SI{1}{min^{-1}}$ resulting in $1440$ shots per day. Assuming a day of data collection per angular step and a required rate of signal photons $R_\gamma=\SI{1}{day}^{-1}$, the projected bounds for this system are
\begin{widetext}
\begin{equation}
    \label{Eq:ProjectedBoundsOptical}
    \gagg\geq\SI{3.5e-7}{GeV^{-1}}\left(\frac{\SI{1.5}{kJ}}{E_1}\right)^\frac{1}{4}\left(\frac{\SI{1.5}{kJ}}{E_2}\right)^\frac{1}{2}\left(\frac{d}{\SI{10}{cm}}\right)^\frac{1}{4}\left(\sqrt{1-\left(\frac{m_a}{\SI{3.08}{eV}}\right)^2}\right)^{\frac{1}{4}}\left(\frac{\SI{3.08}{eV}}{m_a}\right)^2\left(\frac{R_\gamma}{\SI{}{day}^{-1}}\right)^\frac{1}{4},
\end{equation}
\end{widetext}
where we have taken $\ell=\ell'=\tau=\sqrt{d/k_a}$ and quote the bounds for the maximal mass, $m_a\sim\SI{3.08}{eV}$, that can be reached with this setup, obtained by requiring the two beams to be $1$° off the counter-propagation direction.
The testable parameter space is shown in red in figure \ref{fig:Reach} with the dashed red line the projection obtained assuming $E_1=E_2=\SI{15}{kJ}$. Such increase in laser energy may be within reach by the next generation of high-power laser systems. 
The lower cut-off in mass assumes a minimal angle of collision $\alpha=1$°. 
In principle, we can extend the exclusion region to lower masses by exploiting the collision of two photons in a converging beam geometry, at arbitrarily small angles similar to what shown in Ref.~\cite{Nobuhiro:2020fub}. However, in that case, the predicted bounds will fall below those already excluded by PVLAS, and they will not probe any new parameter space. Additional increase in the mass range of predicted bounds, also shown in figure \ref{fig:Reach}, exploits the use of frequency doubled beams. 
In estimating these bounds we have assumed a $10\%$ energy loss for frequency doubling, but such assumption only affects the projected bounds weakly.

To extend the exclusion bounds to even higher axion masses we consider exchanging one of the drive beams with an x-ray free electron laser (XFEL). The European XFEL operates at $\omega_1=\SI{1}{keV}$ with a pulse length $\tau=\SI{100}{fs}$ and energy per pulse of $E_1=\SI{0.5}{ mJ}$. The shorter pulse length limits the interaction region to a cube of side $\tau$ and the resulting bounds are shown in purple in figure \ref{fig:Reach} for the same distance $d=\SI{10}{cm}$. The stimulating laser is a copy of the optical beam to ensure a favourable scaling with the large energy available with such lasers ($E_2=E_s=\SI{1.5}{kJ}$). The right region extends the mass range considerably because the frequency of the European XFEL may be tuned up to $\omega=\SI{25}{keV}$. In principle one can go ahead and exchange all optical beams for XFEL ones, however, due to the decrease in total power this strategy quickly becomes sub-optimal.
In drawing the exclusion regions as continuous areas we made the same set of assumptions as was already discussed in Ref.~\cite{Beyer:2020dag}. In a real laser system the spectral width ensures a width in axion masses we test at each angle. We therefore choose the angular step size such that the excluded region is covered continuously. This is possible in $\sim 30$ steps if we assume a minimal collision angle and minimal step-size of $1$°.

We conclude that the present scheme is capable of producing a competitive $N^3$ scaling with the photon number, and it can access an axion parameter space currently unexplored by laboratory experiments. For higher axion masses, we find that the collision of an optical high power pulse with a x-ray free electron laser produces bounds which still test parameter space formerly not reached by laboratory experiments, however the bounds drop off due to the large decrease of photon numbers in the x-ray beam. Future improvement of laser energy may have the potential to reach the QCD band for $\SI{}{eV}$ masses due to the favourable $N^3$ dependence of the signal photons.

\begin{acknowledgments}
We would like to acknowledge our discussions with Kevin Zhou (Stanford) and Antonino Di Piazza (MPIK) that prompted us working on this topic. 
GG acknowledges partial support from STFC (grant no.~ST/T006277/1). 
The research leading to these results has received funding from AWE plc. British Crown Copyright 2021/AWE.
\end{acknowledgments}

\bibliography{Bibliography}
\bibliographystyle{hunsrt.bst}

\appendix
\begin{widetext}
\section{\label{App-Assym}COUPLING OF DIFFERENT FREQUENCY BEAMS}
With the definition of the geometry \eqref{Eq:GeometryDef} we exploited the symmetry between the two beams present in the collision of two identical (up to polarisation and propagation direction) optical beams. This allowed for simple expressions denoting the dependence on the scattering geometry. When quoting the bounds achievable by the collision of an optical beam with a X-FEL we must drop this assumption. Fixing the geometry to have the axion propagate again in the $\hat{z}$ direction we find
\begin{equation}
    \label{Eq:AssymMomentum1}
    \frac{\mathbf{k}_1}{|\mathbf{k}_1|}=\frac{1}{\sqrt{\omega_1^2+\omega_2^2+2\omega_1\omega_2\cos\alpha}}\left(\omega_2\sqrt{\sin^2\alpha},0,\sqrt{\omega_2^2\cos^2\alpha+\omega_1^2+2\omega_1\omega_2\cos\alpha}\right),
\end{equation}
\begin{equation}
    \label{Eq:AssymElectric1}
    \frac{\boldsymbol{\mathcal{E}}_1}{|\boldsymbol{\mathcal{E}}_1|}=(0,1,0),
\end{equation}
\begin{equation}
    \label{Eq:AssymMagnetic1}
    \frac{\boldsymbol{\mathcal{B}}_1}{|\boldsymbol{\mathcal{B}}_1|}=\frac{-1}{\sqrt{\omega_1^2+\omega_2^2+2\omega_1\omega_2\cos\alpha}}\left(\sqrt{\omega_2^2\cos^2\alpha+\omega_1^2+2\omega_1\omega_2\cos\alpha},0,\omega_2\sqrt{\sin^2\alpha}\right).
\end{equation}
And for the second laser
\begin{equation}
    \label{Eq:AssymMomentum2}
    \frac{\mathbf{k}_2}{|\mathbf{k}_2|}=\frac{1}{\sqrt{\omega_1^2+\omega_2^2+2\omega_1\omega_2\cos\alpha}}\left(\omega_1\sqrt{\sin^2\alpha},0,\sqrt{\omega_1^2\cos^2\alpha+\omega_2^2+2\omega_1\omega_2\cos\alpha}\right),
\end{equation}
\begin{equation}
    \label{Eq:AssymElectric2}
    \frac{\boldsymbol{\mathcal{E}}_2}{|\boldsymbol{\mathcal{E}}_2|}=\frac{1}{\sqrt{\omega_1^2+\omega_2^2+2\omega_1\omega_2\cos\alpha}}\left(\sqrt{\omega_1^2\cos^2\alpha+\omega_2^2+2\omega_1\omega_2\cos\alpha},0,-\omega_2\sqrt{\sin^2\alpha}\right),
\end{equation}
\begin{equation}
    \label{Eq:AssymMagnetic2}
    \frac{\boldsymbol{\mathcal{B}}_2}{|\boldsymbol{\mathcal{B}}_2|}=(0,1,0).
\end{equation}
We then evaluate
\begin{equation}
    \label{Eq:FAssym}
    \mathcal{F}^2=|\boldsymbol{\mathcal{E}}_1|^2|\boldsymbol{\mathcal{E}}_2|^2\left(\frac{\omega_1 \omega_2 \sin ^2\alpha +2 \omega_1 \omega_2 \cos \alpha -\sqrt{(\omega_1 \cos \alpha +\omega_2)^2(\omega_2 \cos \alpha +\omega_1)^2}+\omega_1^2+\omega_2^2}{\omega_1^2+\omega_2^2+2 \omega_1 \omega_2 \cos \alpha }\right)^2,
\end{equation}
and
\begin{equation}
    \label{Eq:j0PerpAssym}
    |\mathbf{j}_0|^2=\omega_a^2 \omega_1^2 |\boldsymbol{\mathcal{E}}_2|^2 \left[\left(1- \frac{k_a}{\omega_a}\sqrt{\frac{(\omega_1 \cos\alpha+\omega_2)^2}{\omega_1^2+\omega_2^2+2 \omega_1 \omega_2 \cos\alpha}}\right)^2+\frac{k_a^2}{\omega_a^2}\frac{(\omega_1 \cos\alpha+\omega_2)^2}{\left(\omega_1^2+\omega_2^2+2 \omega_1 \omega_2 \cos\alpha\right)}\frac{1}{(k_a d)^{2}}\right],
\end{equation}
where again we chose beam $2$ to be the stimulating one. This results in energy of the signal field
\begin{equation}
    \label{Eq:EnergAssym}
    E=\frac{\gagg^4}{256\pi^2} \frac{\ell^2}{d^2}\omega_a^2E_1E_2^2\frac{\mathcal{F}^2}{|\boldsymbol{\mathcal{E}}_1|^2|\boldsymbol{\mathcal{E}}_2|^2}\frac{|\mathbf{j}_0|^2}{\omega_a^2 \omega_1^2 |\boldsymbol{\mathcal{E}}_2|^2 },
\end{equation}
from which we may trivially find the bounds on $\gagg$ as indicated by the purple region in Figure \ref{fig:Reach}.
\end{widetext}

\end{document}